\documentclass[final,3p,times]{elsarticle}

\usepackage{graphicx}
\usepackage{amsmath}
\usepackage{ulem}
\usepackage{float}
\usepackage{comment}

\usepackage{color}

\journal{Physica A: Statistical Mechanics and its Applications}

\begin{document}

\begin{frontmatter}

\title{The correlated cluster mean-field approach to the frustrated Ising model on the honeycomb lattice}


\author[ufms]{Carlos H. D. Batista}
\author[ufsm]{M. Schmidt}
\author[ufms]{F. M. Zimmer}
\address[ufms]{Instituto de F\'isica, Universidade Federal de Mato Grosso do Sul,  79070-900 Campo Grande, MS, Brazil}
\address[ufsm]{Departamento de F\'isica, Universidade Federal de Santa Maria, 97105-900 Santa Maria, RS, Brazil}

\begin{abstract}
 
We study the $J_1$-$J_2$ Ising model on the honeycomb lattice, considering ferromagnetic interactions between first neighbors ($J_1$) and antiferromagnetic interactions between second neighbors ($J_2$).
Our analysis is based on the correlated cluster mean field theory, which is adapted to incorporate competing interactions, providing estimates for the behavior of magnetization, internal energy, entropy, specific heat, and short-range correlations of the model. 
Our results indicate that the transition temperature of the ferromagnetic-paramagnetic phase transition decreases toward zero as the frustration maximum ($J_2/J_1 = -1/4$) is approached, and the thermodynamic quantities indicate only continuous phase transitions for $-1/4<J_2/J_1 \leq 0$.
The critical temperature and the nature of phase transitions provided by the correlated cluster mean-field method are in excellent agreement with very recent Monte Carlo simulations for the model.
Furthermore, the specific heat exhibits a broad maximum within the PM phase under strong frustration,  suggesting the onset of a correlated paramagnetic state with high entropy content at low temperatures. 
Therefore, our findings support that frustration not only suppresses the ferromagnetic long-range order, but also drives significant changes in the thermodynamics and short-range correlations of the model.


\end{abstract}


\end{frontmatter}


\section{Introduction }

Frustration, which is associated with the inability to satisfy all interactions simultaneously, is a promising source of novel physics and challenging problems in spin systems \cite{Vojta_2018, Powell_2011, JPCM_review_2025}. The conflicting situation that often arises in frustrated magnets can lead to many exotic phenomena, including the onset of residual entropy and unusual thermodynamic behavior \cite{Nordblad_2013, Ramirez_review_1994, Savary_2017}, as well as the suppression of conventional long-range orders \cite{claudine, balents2010spin}. In this context, spin models with competing exchange interactions between first ($J_1$) and second neighbours ($J_2$)  have played a central role in the investigation of frustration effects by allowing the control of the level of frustration through the ratio $J_2/J_1$ \cite{CRPHYS_2025}. For instance, numerous works have been dedicated to the investigation of the phase transitions of the $J_1$ - $J_2$ Ising model on the square lattice \cite{PhysRevE.91.032145, PhysRevE.99.012134, PhysRevE.97.022124, DOSANJOS20081180, PhysRevE.104.024118, PhysRevB.109.104419, PhysRevE.107.034124, PhysRevB.21.1941, PhysRevB.84.174407, PhysRevB.87.144406, PhysRevLett.108.045702, PhysRevB.86.134410, PhysRevB.109.064422, kalz2008phase, PhysRevB.48.3519, Aguilera-Granja_1993, LOPEZSANDOVAL1999437, PhysRevE.91.052123, PhysRevE.108.054124, PhysRevB.104.144429, PhysRevE.111.024109, PTEP_2024}, with significant efforts made to locate a tricritical point in the model.

From the experimental side, the significant progress made in the synthesis and characterization of van der Waals magnets allowed the experimental study of low-dimensional systems \cite{acsnano_vdW}, in which frustration effects can be enhanced. Interestingly, several of these materials form magnetic honeycomb lattices, often exhibiting strong Ising anisotropy.  
For instance, the compound FePS$_3$ hosts a zigzag antiferromagnetic phase that persists even to the monolayer limit \cite{PhysRevB.101.024415, nanolett_feps3}. Other prototypical system is VBr$_3$, which also exhibits an Ising-like antiferromagnetic phase with a field-induced tricritical point \cite{PhysRevB.108.104416}. Despite the strong  anisotropy, a simple Ising model with isotropic interactions between first-neighbour spins is unable to describe the unusual form of antiferromagnetism hosted by both FePS$_3$ and VBr$_3$, suggesting that a microscopic model 
including farther-neighbor interactions is required \cite{PhysRevB.101.024415, PhysRevB.108.104416, PhysRevB.108.014436}. 
%
The $J_1$-$J_2$ Ising model on the honeycomb lattice arises as an interesting attempt to describe the magnetic behavior of frustrated van der Waals compounds with uniaxial anisotropy, revealing that competing interactions can play a central role in honeycomb-lattice magnets. However, this model has historically received less attention than its square-lattice counterpart.

The investigations conducted in the past decade have sparked a debate concerning the phase transitions and thermodynamic properties hosted by this frustrated Ising honeycomb system. 
The presence of ferromagnetic ($J_1>0$) and antiferromagnetic ($J_2<0$) interactions with first and second neighbors, respectively, leads to a ground-state transition from a ferromagnetic (FE) phase to a degenerate state at $J_2/J_1=-1/4$ \cite{MC_Wang-Landau}.   %
%
The nature of the thermal phase transition between FE and paramagnetic (PM) phases has been the main focus of recent studies. In particular, the effective field approach using clusters \cite{BOBAK20162693} indicates the presence of a tricritical point at the FE-PM phase boundary, while the cluster mean-field method \cite{SCHMIDT2021168151} suggests only continuous phase transitions. Monte Carlo simulations using standard Metropolis and parallel tempering approaches confirm the onset of continuous phase transitions for $-0.20 \leq J_2/J_1 \leq 0$ \cite{ZUKOVIC2021127405}. However, the slow convergence to the equilibrium prevented any conclusion concerning the phase transitions for stronger antiferromagnetic couplings between second-neighbors. Recent investigations based on partition function zeros \cite{Gessert2024PartitionFZ} and population annealing Monte Carlo simulations \cite{arxiv_gessert} showed that the transitions remain continuous for $J_2/J_1 = -0.22$ and $-0.23$, respectively, with an intricate scenario arising for $J_2/J_1$ very close to $-1/4$ within both approaches. Furthermore, a scenario of only continuous FE-PM phase transitions is corroborated by recent Wang-Landau simulations \cite{MC_Wang-Landau}, but this numerical analysis was limited to relatively small system sizes. Therefore, further studies are required to confirm the nature of the FE-PM phase transitions hosted by the model, especially close to $J_2/J_1=-1/4$.

In the present work, we study the phase transitions of the $J_1$-$J_2$ Ising model on the honeycomb lattice by employing the correlated cluster mean-field (CCMF) theory \cite{Yamamoto}, which is based on the correlated molecular-field method \cite{PhysRevE.61.6399} within a cluster concept.
This method allows obtaining a more precise estimate for the critical temperature of Ising spin models when compared to the effective field theory and cluster mean-field methods without the high computational cost and intricate aspects of numerical simulations \cite{SCHMIDT2021125884}. 
It is worth noting that the specific heat computed for $J_2/J_1\approx -1/4$  within cluster mean-field theory \cite{SCHMIDT2021168151}, as well as Wang-Landau \cite{MC_Wang-Landau} and population annealing \cite{arxiv_gessert} Monte Carlo simulations, exhibits a round maximum for temperatures above the FE-PM transition, a feature usually attributed to frustration effects \cite{Savary_2017}. Motivated by this behavior, we also investigate thermodynamic quantities of the model, including specific heat, entropy, and internal energy. We remark that the CCMF approach provides an accurate description of the specific heat and entropy of the strongly frustrated Ising kagome antiferromagnetic \cite{Schmidt_2017}, being a suitable framework to investigate frustration effects on thermodynamic quantities. Finally, we dedicate our attention to the effects of frustration on the short-range spin-spin correlations of the model, a subject overlooked in previous studies.

\par This article is structured as follows. The model discussed within the CCMF approach is presented in Sec. \ref{secmodel}. Section \ref{secresults} is reserved for the numerical results of the thermodynamic quantities under different conditions, followed by a discussion of their properties. The conclusion as well as future research perspectives are presented in Sec. \ref{secconclusion}.





\section{Model}\label{secmodel}

We consider the $J_1 - J_2$ Ising model on the honeycomb lattice structure, described by the Hamiltonian 
\begin{equation}\label{model}
{\cal H} = -J_{1}\sum_{\langle i,j \rangle} \sigma_{i}\sigma_{j} - J_{2}\sum_{\langle\langle i,k \rangle\rangle} \sigma_{i}\sigma_{k},
\end{equation}
where $\sigma_i=\pm 1$ represents the Ising spin variables at the site $i$. The summations $\langle i,j \rangle$ and $\langle\langle i,k \rangle\rangle$ run over nearest and next-nearest neighbor sites, respectively. We assume ferromagnetic exchange interactions ($J_1>0$) between the nearest-neighbors and antiferromagnetic exchange interactions ($J_2\leq0$) between the next-nearest-neighbor sites.
The partition function of this model is known exactly for the limit case $J_2=0$ \cite{Houtappel1950}, in which a FE-PM phase transition takes place at the critical temperature $k_B T_c/J_1 = 2/\ln{(2+\sqrt{3})} \approx 1.519$ \cite{Houtappel1950}. At absolute zero, the ground-state energy per spin of the FE phase is given by $u_\textrm{FE} = - 3J_1/2 -3J_2$. Therefore, the antiferromagnetic next-nearest-neighbor couplings compete with the ferromagnetic nearest-neighbor couplings, increasing the energy of the FE phase. At $J_2/J_1=-1/4$, a transition occurs in the ground-state to a highly degenerate stripe state, whose energy is $U_{\textrm{ST}}=-J_1/2 +J_2$ \cite{Katsura1986}.  Therefore, a suppression of the FE ordering is expected to take place at finite temperatures when $J_2/J_1 \to -1/4$.

\begin{figure*}[t]
\centering
\includegraphics[width=0.85\textwidth]{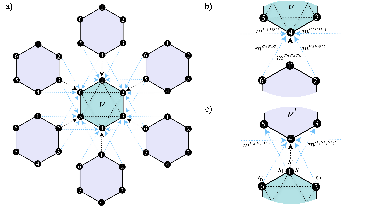}
\caption{Schematic representation of the honeycomb lattice divided into clusters, where solid lines indicate interactions between first neighbors ($J_1$) and dashed lines represent interactions between second neighbors ($J_2$). Arrows indicate the mean fields acting on the central cluster $\nu$. In panel (b), the five effective fields acting on site 4 of the central cluster $\nu$ are shown, where the dashed black arrow indicates the effective field arising from a first-neighbor interaction, and the dashed blue arrows indicate the effective fields from a second-neighbor interaction. Panel (c) shows the neighbor cluster $\nu'$ with the mean fields from cluster $\nu$ replaced by parameters $s_6$, $s_1$, and $s_2$.}
\label{hexagonalcentral}
\end{figure*}

To study the model defined in Eq. (\ref{model}), we adapt the correlated cluster mean-field (CCMF) theory \cite{Yamamoto}. 
In this approach, the spin lattice is divided into identical clusters of $n_s$ spins, ensuring that the resulting pattern of repeating clusters contains all the sites of the original lattice. 
The cluster division allows the identification of two sets of interactions: intracluster and intercluster interactions. 
The intracluster interactions are fully preserved, while the intercluster interactions are replaced by correlated effective fields to be self-consistently determined. 
This procedure provides an expression of the model in a clustered structure, in which 
the effective  single-cluster model is given by
\begin{equation}\label{eq2}
{\cal H}_\nu=- J_{1} \sum_{\langle i,j \rangle } \sigma_{i} \sigma_{j} - J_{2} \sum_{\langle\langle i,k \rangle\rangle } \sigma_{i} \sigma_{k} - \sum_{i\in \bar{\nu}} \sigma_{i} h_{i}^{\text{eff}} (\{\sigma\}),
\end{equation}
where $h_{i}^{\text{eff}}(\{\sigma\})$ represents the effective field acting on the spin at site $i$ on the cluster boundary $\bar{\nu}$. This effective field depends on the spin configurations of the cluster itself, which interact with the neighboring cluster. 
Therefore, the number of possible effective fields depends on the lattice geometry and the adopted cluster structure. In our case, we consider the lattice divided into honeycomb clusters of six sites, as illustrated in Fig. \ref{hexagonalcentral}(a).

For the cluster adopted, each effective field acting on each site $i$ must incorporate the five interactions between the clusters as mean fields: one from nearest-neighbor sites and four from next-nearest-neighbor sites, all originating from sites belonging to neighboring clusters.

To illustrate how the mean fields are defined within the CCMF method, let us consider two neighboring clusters, $\nu$ and $\nu^{'}$ (see Fig. \ref{hexagonalcentral}(b)). 
The set of mean fields depends on the spin state of the three sites of cluster $\nu$ that interact with the spins of cluster $\nu^{'}$. Therefore, each of the eight possible spin configurations of these three spins corresponds to a distinct mean field. Explicitly, the interactions between the clusters are replaced by mean-field terms $m^{+++}$, $m^{++-}$, $m^{+-+}$, $m^{-++}$, $m^{--+}$, $m^{-+-}$, $m^{+--}$, and $m^{---}$, where subscript represents the possible spin configurations. For instance, when the spins $\sigma_3,\sigma_4,\sigma_5$ are in the state $|\uparrow\downarrow\uparrow\rangle$, the chosen mean field is $m^{+-+}$, and so on. 
It is worth noting that in the absence of second neighbor interactions, the CCMF approach is reduced to two independent mean-fields \cite{Yamamoto}. To deal with the present model, we adapt this theoretical framework, incorporating effects of frustration introduced by $J_2$.
In this adapted CCMF, the interactions between clusters $\nu$ and $\nu^{'}$ can be expressed in terms of the mean field as
$[(\sigma_{3{'}}+\sigma_{5{'}})J_2+\sigma_4{'}(J_1+2J_2)]m^{\sigma_3{'},\sigma_4{'},\sigma_5{'}}$, where the superscript $\sigma_3{'},\sigma_4{'},\sigma_5{'}$ denotes the spin projection of sites 3{'}, 4{'} and 5{'}, with $\sigma=\uparrow$ or $\sigma=\downarrow$ corresponding to $+$ or $-$, respectively. 
By following this procedure for the other boundary sites, we can express the effective Hamiltonian as
%
\begin{equation}\begin{split}\label{heff10}
{\cal H}_\nu=- J_{1} \sum_{\langle i,j \rangle } \sigma_{i} \sigma_{j} - J_{2} \sum_{\langle\langle i,k \rangle\rangle } \sigma_{i} \sigma_{k} 
- \sum_{i\in \bar{\nu}} [J_2(\sigma_{n}+\sigma_{\bar{n}})+(J_1+2J_2)\sigma_i ]m^{\sigma_{n}\sigma_i\sigma_{\bar{n}}},
\end{split}
\end{equation}
where $n$ and $\bar{n}$ are intracluster nearest neighbors of $i$.

To obtain the mean fields, we consider the neighboring connected clusters, say $\nu'$. From this, we compute the expected value of nearest-spin of cluster $\nu$, $m^{s_6 s_1 s_2}=\langle\sigma_{4^{'}}\rangle_{{\cal H}_{\nu^{'}}(s_6,s_1,s_2)}$, in which the thermal average  is taken with respect to the auxiliary cluster model ${\cal H}_{\nu^{'}}(s_6,s_1,s_2)$.  This model considers the same cluster structure as the effective Hamiltonian with the effective fields replacing the intercluster interactions except for the direct couplings between spins of clusters $\nu$ and $\nu^{'}$. These interactions are evaluated assuming that the spins of the neighboring cluster $\nu$ have magnetic moments given by the values $s_6,s_1,s_2$.  The mean fields are then self-consistently computed for each of the eight possible states of the three boundary spin sites:
\begin{equation}
m^{s_6 s_1 s_2} = 
\frac{\text{Tr} \left( \sigma_{4'} \mbox{e}^{-\beta {\cal H}_{\nu'}(s_6 s_1 s_2)} \right)}{\text{Tr} \left( \mbox{e}^{-\beta {\cal H}_{\nu'}(s_6 s_1 s_2)} \right)},
\label{eq:meanfields}
\end{equation}
where $\beta$ = $1/k_BT$ ($k_B$ is the Boltzmann constant and $T$ is the temperature), Tr represents the trace over cluster spin states, and

\begin{equation}
\begin{split}
{\cal H}_{\nu'}(s_6,s_1,s_2)=- J_{1} \sum_{\langle i,j \rangle } \sigma_{i} \sigma_{j} - J_{2} \sum_{\langle\langle i,k \rangle\rangle } \sigma_{i} \sigma_{k} -
\sum_{i\neq\{ 3',4{'},5{'}\}} [(J_1+2J_2)\sigma_i+J_2(\sigma_{n}+\sigma_{\bar{n}})]m^{\sigma_{n}\sigma_i\sigma_{\bar{n}}}
-J_2(\sigma_{3{'}}+\sigma_{5{'}})s_1 \\ -J_1\sigma_{4{'}}s_1  - 
(J_1+2J_2)[\sigma_{3{'}} m^{\sigma_{2{'}}\sigma_{3{'}}\sigma_{4{'}}}
+\sigma_{5{'}} m^{\sigma_{4{'}}\sigma_{5{'}}\sigma_{6{'}}}]
-J_2\sigma_{4{'}} [ m^{\sigma_{2{'}}\sigma_{3{'}}\sigma_{4{'}}}
+ m^{\sigma_{4{'}}\sigma_{5{'}}\sigma_{6{'}}} + s_2 + s_6]
.
\label{eq:HAMILTONIANOAUXILIAR}
\end{split}
\end{equation}

With the mean field evaluated, we can return to the effective Hamiltonian to obtain, for instance, the magnetization per spin $(m)$ 
\begin{align}
m = \sum_{i}\langle \sigma_{i} \rangle/n_s = \sum_{i}\frac{ \text{Tr}(\sigma_{i} e^{-\beta  {\cal H}_V})}{n_s\text{Tr} (e^{-\beta  {\cal H}_V} )}. 
   \label{eq:magnetizaçãospin1}
\end{align}
In the absence of an external magnetic field, magnetization can be used as an order parameter to identify the presence of the ferromagnetic phase and to characterize the phase transition to the paramagnetic phase.
Alternatively, phase transitions are often described in terms of other thermodynamic quantities, such as entropy and specific heat. These quantities can be obtained straightforwardly if the free energy of the system is known. However, a closed form for the free energy is not provided in the original formulation of the CCMF theory \cite{Yamamoto}. In a mean field spirit, one can propose that the free energy per cluster within the CCMF framework is given by 
$f=-k_B T \ln \text{Tr} \, e^{-{\beta\cal H}_\nu}+\gamma_{CCMF}$, where $\gamma_{CCMF}$ is a correction. We note that an analogous form for the free energy is obtained within the variational cluster mean-field approach, where the correction is associated with a perturbative term that is a function of local magnetizations \cite{PhysRevE.98.022123}. 
Similarly, the internal energy per cluster can be written as a sum of the thermodynamic average of the effective Hamiltonian and the mean-field correction \cite{SCHMIDT2021168151}. Analogously, within the CCMF approach, the internal energy per cluster would be given by $U= \text{Tr}{\cal H}_{\nu}e^{-{\beta\cal H}_\nu}/\text{Tr}e^{-{\beta\cal H}_\nu}+\gamma_{CCMF}$. In the present work, our lack of knowledge about the mean-field corrections prevents us from obtaining both thermodynamic quantities, $f$ and $U$, from the equations above. However, given that the proposed forms for the free energy and the internal energy are appropriate, the entropy per cluster, which can be written in terms of the difference between $U$ and $f$, is independent of the correction. Therefore, we compute the entropy per spin as
\begin{equation}
S=\frac{U-f}{T n_s} = \frac{1}{n_s T }\left( \frac{\text{Tr}{\cal H}_{\nu}e^{-{\beta\cal H}_\nu}}{ \text{Tr}e^{-{\beta\cal H}_\nu}} + k_B T \ln \text{Tr} \, e^{-{\beta\cal H}_\nu} \right).
\end{equation}

Within the present self-consistent approach, the internal energy can be estimated by using the spin-spin correlations. We compute the internal energy per spin as $u= - 3 J_1 C_1/2 - 3 J_2 C_2$, where $C_1=\sum_{\langle i,j \rangle} \langle \sigma_i \sigma_j\rangle_{H_\nu}/6$ and $C_2=\sum_{\langle\langle i,k \rangle\rangle} \langle \sigma_i \sigma_k\rangle_{H_\nu}/6$ are correlations between nearest- and next-nearest-neighbors, respectively. Here, only intracluster interactions are considered. We also investigate the specific heat $C_v= \frac{d u}{dT}$, evaluated from a numerical derivative of the proposed internal energy.


\section{Results}\label{secresults}

\begin{figure}[b]
\centering
\includegraphics[width=0.9\columnwidth]{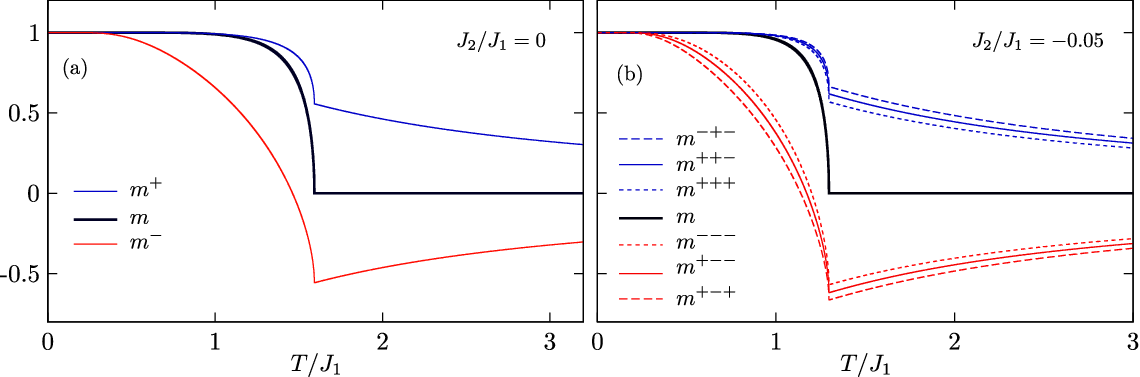}
\caption{Mean fields and magnetization as functions of temperature for (a) $J_2/J_1 = 0$ and (b) $J_2/J_1 = -0.05$. The magnetization is represented by the thick black line, while the thin lines correspond to the mean fields.}
\label{mean-fields}
\end{figure}

The numerical results are obtained by solving self-consistently the set of eight mean-field equations (given by Eq. (\ref{eq:meanfields}) for each of the possible spin states for three sites)  with the auxiliary Hamiltonian defined in Eq. (\ref{eq:HAMILTONIANOAUXILIAR}).
Exact numeration is used to compute the auxiliary Hamiltonian.
Once the mean fields are determined, the Hamiltonian given by Eq. (\ref{heff10}) is used to evaluate thermodynamic quantities, such as magnetization, spin-spin correlations, internal energy, specific heat and entropy.
We use the exchange interaction $J_1$ as the energy unit and, for numerical purposes, set $k_B=1$.

Figure \ref{mean-fields} shows the temperature dependence of the mean fields and magnetization per spin for $J_2/J_1=0$ and $J_2/J_1=-0.05$. At high temperatures, the mean fields exhibit symmetric finite values around zero, and the magnetization is zero, indicating the PM phase. 
As the temperature decreases, the mean fields lose symmetry at the critical temperature $T_c$, resulting in a finite magnetization and the system becoming FE.
We remark that there is an important difference between the mean fields in the absence of next-nearest-neighbor couplings and for a finite $J_2/J_1$.
From the eight possible mean-field configurations representing the intercluster interactions, there are only two independent parameters that  are required to account for the single intercluster interactions in the case of $J_2=0$, as shown in Fig. \ref{mean-fields}(a). This means that $m^{+}\equiv m^{+++}=m^{-++}=m^{-+-}=m^{++-}$ and $m^{-}\equiv m^{---}=m^{--+}=m^{+--}=m^{+-+}$, which recovers the findings presented in Ref \cite{Yamamoto} with $T_c/J_1=1.593$ for the Ising model on the honeycomb lattice studied with the CCMF method. When $|J_2|>0$, the symmetry of the cluster allows us to obtain two equalities: $m^{++-}=m^{-++}$ and $=m^{+--}=m^{--+}$. In this case, six independent mean field parameters should be computed to incorporate the intercluster interactions within the CCMF approach. This scenario is illustrated in Fig. \ref{mean-fields}(b) for $J_2/J_1=-0.05$, in which a critical temperature $T_c/J_1=1.297$ is indicated by the behavior of the magnetization per site.
This points out that the magnitude of the ratio $|J_2/J_1|$ affects not only the mean fields but also the location of $T_c/J_1$.

\begin{figure*}[t]
\centering
\includegraphics[width=0.99\textwidth]{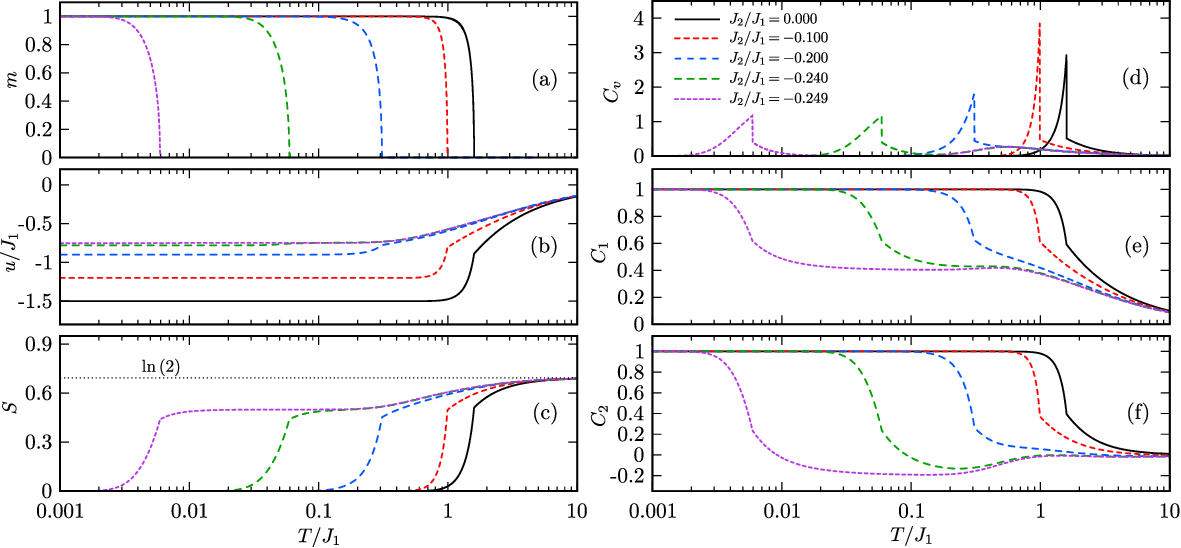}
\caption{(a) Magnetization per spin, (b) internal energy per spin, (c) entropy per spin, (d) specific heat (e) nearest-neighbor correlations, and (f) next-nearest-neighbor correlations as functions of temperature (in logarithmic scale) for several strengths of the next-nearest-neighbor interactions.
}
\label{Fig_mag}
\end{figure*}

Figure \ref{Fig_mag}(a) shows magnetization curves for several strengths of second neighbor couplings, evidencing that the critical temperature  $T_c/J_1$ gradually decreases as the antiferromagnetic interactions between next-nearest neighbours are enhanced.
It is important to note that the magnetization exhibits continuous behavior (i.e., it decreases continuously from 1 to 0 at the critical temperature)  for all the curves depicted in Fig. \ref{Fig_mag}(a), suggesting continuous phase transitions between the PM and FE phases.

\begin{figure}[t]
\centering
\includegraphics[width=0.99\columnwidth]{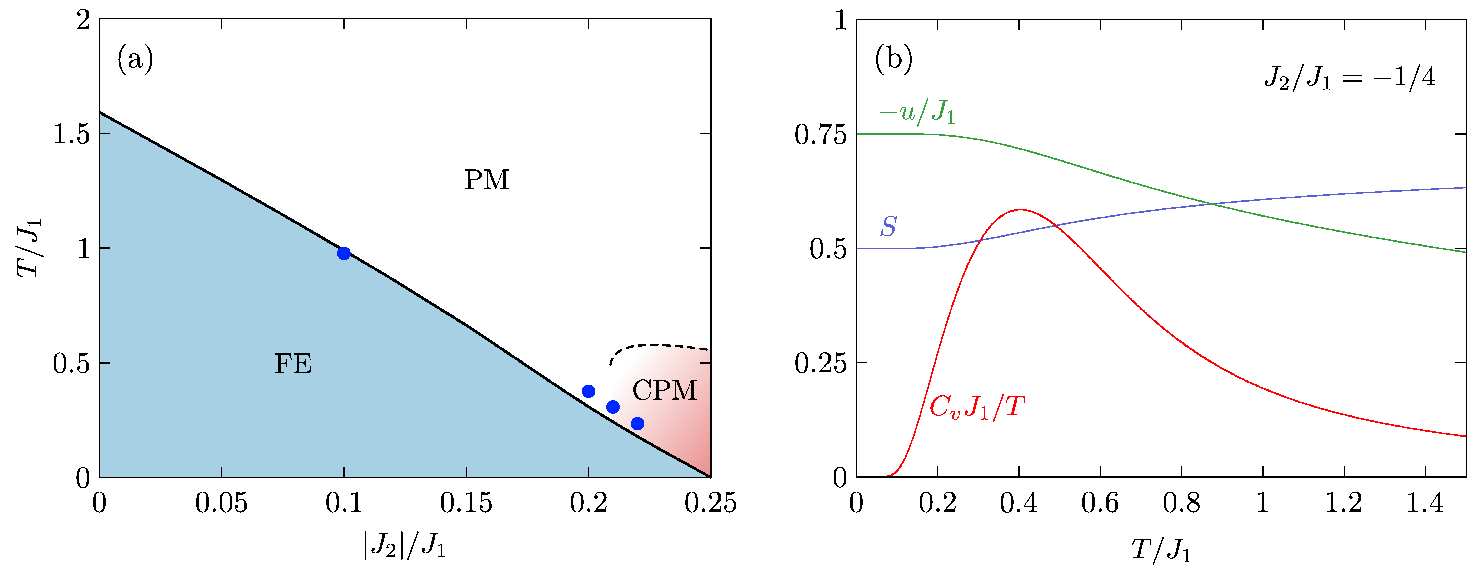}
\caption{ (a) Phase diagram of temperature versus coupling. The solid line corresponds to the FE-PM phase boundary, while the dashed line indicates the round maximum of the specific heat as a function of temperature within the  CCMF method. For temperatures between $T_c$ and the specific heat round maximum, negative spin-spin correlations between next-nearest neighbors and a plateau-like behavior of entropy suggest the formation of a correlated paramagnetic (CPM) state. Only continuous FE-PM phase transitions are obtained within the present approach. Filled blue circles are results from scaling analysis of partition function zeros reported in Ref. \cite{Gessert2024PartitionFZ}. (b) Internal energy, entropy and specific heat divided by temperature as functions of temperature in the scenario of maximum frustration $J_2/J_1=-1/4$. }
\label{Temperaturacritica1}
\end{figure}

To help our analysis of the thermodynamic properties, Fig. \ref{Fig_mag} also displays the behavior of the internal energy, specific heat, entropy, and short-range spin-spin correlations.
For instance, the internal energy exhibits continuous behavior, marked by an inflection point at the critical temperature, which is characteristic of a second-order phase transition (see Fig. \ref{Fig_mag}(b)). It smoothly approaches zero at high temperatures, regardless of the interaction strength. On the other hand, the internal energy exhibits a strong dependence on the exchange interactions at low temperatures, increasing when larger antiferromagnetic interactions between next-nearest neighbors are considered. This increase in the internal energy can be seen as a consequence of the competition between $J_1$ and $J_2$, leading to a FE phase with higher internal energy and enhanced sensitivity to thermal fluctuations. We also note that the ground-state internal energy estimated within the CCMF approach reproduces the exact value $u_{FE}/J_1= -1.5-3J_2/J_1$ for the FE phase. 

An interesting feature arises from the entropy behavior displayed in Fig. \ref{Fig_mag}(c). When thermal fluctuations dominate the system, the entropy per spin approaches $\ln(2)$ in units of $k_B$, independently of the value of $J_2/J_1$. This reflects the high-temperature PM phase, characterized by equal probabilities for both up and down spin states. 
At low temperatures, the ordered FE state exhibits zero entropy.
The entropy presents a typical behavior of a continuous FE–PM phase transition, with a marked continuous drop as temperature is reduced just below $T_c$ for all values of $J_2/J_1$ analyzed.
It is worth noting that, as $J_2/J_1$ approaches -0.25, a high-intensity entropy plateau can be found in the low-temperature PM phase for temperatures just above the transition to the FE phase. 
This behavior may reflect the frustration induced by the competition between  exchange interactions,
resulting in a highly entropic low-temperature PM phase. 

The specific heat reflects the hallmark of a continuous phase transition within a mean-field-like approach, displaying a discontinuity at the critical temperature, as shown in Fig.\ref{Fig_mag}(d). In addition,  a broad specific-heat maximum can be observed for temperatures above $T_c$, as shown for $J_2/J_1=-0.240$ and $J_2/J_1=-0.249$.   This $C_v$ behavior is comparable to that one reported in Ref. \cite{MC_Wang-Landau} for  $J_2/J_1=-0.240$, where a round maximum appears at temperatures above the discontinuity that indicates  $T_c$. Analogous findings were also reported within a cluster mean-field investigation \cite{SCHMIDT2021168151}. Therefore, this round maximum of $C_v(T)$ near the frustration maximum can be expected to be a property of the $J_1-J_2$ Ising model on the honeycomb lattice instead of an artifact of the CCMF approximation.
The temperature range between $T_c$ and the round maximum of $C_v$ is characterized by internal energy and entropy exhibiting a weak dependence on temperature as shown in Fig. \ref{Fig_mag}. To further reason about the properties of the model on this special range of temperature, we present the short-range correlations of the model in panels (e) and (f) of Fig. \ref{Fig_mag}. In general, spin-spin correlations exhibit a positive value for temperatures below the critical temperature, going to one as the absolute zero temperature is approached, which is a consequence of the ferromagnetic ordering. We also note that correlations approach zero as temperature is increased above the ordering temperature, which is expected in a high-temperature PM state. However, correlations $C_1$ and $C_2$ exhibit a plateau-like feature for temperatures between the specific heat maximum and $T_c$, which is particularly noticeable for $J_2/J_1=-0.249$. In addition, next-nearest-neighbor correlations $C_2$ exhibit a negative value in this range of temperature, indicating an enhanced effect of the antiferromagnetic couplings $J_2$. 

To explore the phase transitions indicated in Figs. \ref{mean-fields} and \ref{Fig_mag}, we present in Fig. \ref{Temperaturacritica1}(a) the phase diagram obtained for the model in the coupling temperature plane.
The FE order observed at lower temperatures undergoes a transition to the PM phase at sufficiently high temperatures, which depends on the strength of the exchange interactions. When the strength of $J_2$ increases, $T_c$ indicated by the solid line gradually decreases, reaching zero at $|J_2|/J_1=0.250$. For $|J_2|/J_1>0.250$, the FE order is no longer found. 
Figure \ref{Temperaturacritica1}(a) also exhibits results from a scaling analysis of partition function zeros reported in Ref.~\cite{Gessert2024PartitionFZ} (blue circles).
The $T_c/J_1$ curve obtained with the CCMF method agrees well with the critical temperatures from the partition function zeros analysis, which, in turn, show excellent agreement with the $T_c$ from the Wang–Landau algorithm \cite{MC_Wang-Landau}.  In particular, for $J_2/J_1=-0.10$, the CCMF method indicates $T_c/J_1 \approx 0.99 $ while the analysis in Ref.~\cite{Gessert2024PartitionFZ} suggests $T_c/J_1 \approx 0.98$.
It is noteworthy that the precise determination of the nature and location of the PM-FE transition was not achieved through the  parallel tempering algorithm of Ref. \cite{ZUKOVIC2021127405} in the highly frustrated region of $0.200<|J_2|/J_1<0.250$.  
Furthermore, our results indicate a continuous PM–FE phase transition (as discussed in detail with the support of Fig. \ref{Fig_mag}) without evidence of a tricritical point, in perfect agreement with the Wang–Landau algorithm \cite{MC_Wang-Landau}, partition function zeros \cite{Gessert2024PartitionFZ}, and the cluster mean-field calculations reported in Ref.~\cite{SCHMIDT2021168151}.

The unusual behaviors of entropy, internal energy, and correlations, suggest that the PM state in the range of temperatures between the specific heat round maximum and $T_c$ differs significantly from the usual high-temperature PM phase. We propose that a correlated paramagnetic (CPM) state, originating from the strong competition between exchange interactions, takes place in this range of temperature. Figure \ref{Temperaturacritica1}(a) indicates the range of parameters in which the CPM state occurs, with the dashed lines indicating the temperature of the round maximum of $C_v(t)$. It is also worth noting that the presence of a round maximum in the specific heat within the PM phase has been reported in several frustrated spin systems. We remark that in the present system, a small entropy release occurs for temperatures below the specific heat maximum (see Figs. \ref{Fig_mag}(c)-(d)). More importantly, the entropy exhibits a finite plateau in its temperature dependence within the CPM state. 
It means that the system presents a high entropy at low temperatures, suggesting that frustration plays a major role in the onset of the CPM state. 
Particularly, Fig. \ref{Temperaturacritica1}(b) shows the temperature dependence of the energy, entropy, and specific heat for the frustration maximum $J_2/J_1=-0.25$, where no finite-temperature phase transition takes place. In this case, we find a residual entropy $S(T\rightarrow 0)=S_{res}=0.499$, providing an estimate of the ground-state entropy in a highly frustrated regime. In addition, the specific heat divided by temperature exhibits a broad maximum at a finite temperature, indicating a gradual release of entropy. Finally, the internal energy reaches $u=-0.75J_1$ at the ground state, as expected for the highly frustrated state at $J_2/J_1=-1/4$ \cite{Katsura1986}.

\section{Conclusion}\label{secconclusion}

We have studied the frustrated $J_1$-$J_2$ Ising model on the honeycomb lattice by employing the CCMF theory. Our findings for the thermodynamics of the model indicate the onset of continuous FE-PM phase transitions for $-0.25\leq J_2/J_1<0$, in agreement with results from cluster mean-field theory \cite{SCHMIDT2021168151}, partition function zeros \cite{Gessert2024PartitionFZ}, and very recent Monte Carlo simulations based on the Wang-Landau algorithm \cite{MC_Wang-Landau}. The present CCMF approach also suggests the onset of a round maximum of the specific heat within the PM phase for $-0.25<J_2/J_1<0.21$. For temperatures between this maximum and the FE-PM phase transition, entropy, internal energy, and short-range correlations exhibit a weak temperature dependence, forming a plateau-like structure. In addition, negative correlations between next-nearest-neighbors indicate the onset of frustration effects introduced by $J_2$, suggesting the formation of a CPM regime.

Therefore, our CCMF investigation corroborates the findings from state-of-the-art methods for the FE-PM phase transition, also revealing the presence of strong frustration effects, including residual entropy for $J_2/J_1=-0.25$ and a round maximum of $C_v$ near the frustration maximum. It is worth noting that the CCMF method requires a smaller computational effort when compared to Monte Carlo simulations, which is particularly relevant near the highly frustrated regime, where the implementation of the standard Metropolis algorithm is very intricate \cite{ZUKOVIC2021127405}.
Although the presence of a specific heat maximum within the PM phase was obtained within the cluster mean-field method \cite{SCHMIDT2021168151} and Monte Carlo simulations \cite{MC_Wang-Landau}, the behavior of correlations in this model is a subject that deserves further attention. An interesting question concerns the decay of correlations with distance in the proposed CPM state. We hope that the present investigation motivates further studies of this issue. Another interesting subject concerns the onset of phase transitions for $J_2/J_1<-0.25$, a regime in which tricriticality has been suggested to exist \cite{MC_Wang-Landau}. However, the high degeneracy associated with the low-temperature state found in this regime requires further adaptation of the present CCMF approach, which is left for a future investigation.

\section{Acknowledgments}
This work was supported by the Brazilian funding agencies Conselho Nacional de Desenvolvimento Cient\'ifico e Tecnol\'ogico (CNPq), and Coordena\c{c}\~ao de Aperfei\c{c}oamento de Pessoal de N\'ivel Superior (Capes). FMZ also acknowledge support from the Fundação de Apoio ao Desenvolvimento do Ensino, Ciência e Tecnologia do Estado de Mato Grosso do Sul (Fundect). 
\appendix

\bibliography{reference}

\end{document}